\documentclass[11pt]{article}

\usepackage[latin1]{inputenc}
\usepackage[T1]{fontenc}
\usepackage{amsmath,amsfonts,amssymb,bm,epsf}
\usepackage{pslatex}
\usepackage{graphicx,color}
\usepackage[a4paper,left=2.5cm,right=2.5cm,top=3cm,bottom=3cm]{geometry}
\usepackage{fancyhdr}

\usepackage{multicol}

\newcommand{\xx}{{\bf x}}
\newcommand{\yy}{{\bf y}}

\newcommand{\RR}{{\mathbb R}}

\newcommand{\1}{{\bf 1}}

\makeatletter

\newcommand{\titlerunning}[1]{\def\@titlerunning{#1}}
\newcommand{\authorrunning}[1]{\def\@authorrunning{#1}}
\newcommand{\address}[1]{\def\@address{#1}}
\fancypagestyle{plain}
\AtBeginDocument{\lhead{\textit{\@authorrunning}}\chead{}\rhead{\textit{\@titlerunning}}}

\def\@abstract{\normalsize\textbf{Abstract. }} \def\end@abstract{\par}
\def\@keywords{\normalsize\textbf{Keywords. }} \def\end@keywords{\par}
\newtoks\abstract \newtoks\keywords

\def\acknowledgments{\@startsection{subparagraph}{6}{\z@}{-24pt plus 6pt minus 1pt}{-.5em}
{\bfseries}*{Acknowledgments. \/}}

\renewcommand{\@maketitle}{
\newsavebox{\foo}
\savebox{\foo}{
\begin{minipage}[t]{15.7cm}
	\par\hspace{24.75em}
	
\begin{center}
		\Large\bfseries\@title
		\par\vspace{1em}
		\normalfont\normalsize\noindent\@author
		\end{center}

	\par\vspace{1em}\itshape\small
	\@address
	\par\vspace{1em}\noindent\rule{\linewidth}{.5pt}
	\par\vspace{1em}
	\begin{@abstract}\the\abstract\end{@abstract}
	\par\vspace{1em}
	\begin{@keywords}\the\keywords\end{@keywords}
	\par\vspace{1em}\noindent\rule{\linewidth}{.5pt}
\end{minipage}}
\par\vspace{1em}\noindent\usebox{\foo}
\vspace{1em}}
\renewenvironment{thebibliography}[1]
     {\section*{\large\refname}
      \small\@mkboth{\MakeUppercase\refname}{\MakeUppercase\refname}
      \list{\@biblabel{\@arabic\c@enumiv}}
           {\settowidth\labelwidth{\@biblabel{#1}}
            \leftmargin\labelwidth
            \advance\leftmargin\labelsep
            \@openbib@code
            \usecounter{enumiv}%
            \let\p@enumiv\@empty
            \renewcommand\theenumiv{\@arabic\c@enumiv}}
      \sloppy
      \clubpenalty4000
      \@clubpenalty \clubpenalty
      \widowpenalty4000%
      \sfcode`\.\@m}
     {\def\@noitemerr
       {\@latex@warning{Empty `thebibliography' environment}}
      \endlist}
\makeatother

\title{A simulated annealing procedure based on the ABC Shadow algorithm for statistical inference of point processes}
\titlerunning{A simulated annealing procedure based on the ABC Shadow algorithm}

\author{R. S. Stoica$^{1,*}$, M. Deaconu$^{1,2}$ and L. Hurtado$^3$}
\authorrunning{R. S. Stoica \emph{et al.}}

\address{
$^1$ Universit\'e de Lorraine, CNRS, IECL, F-54000, Nancy, France ; radu-stefan.stoica@univ-lorraine.fr\\
$^2$ Universit\'e de Lorraine, CNRS, Inria, IECL, F-54000, Nancy, France ; madalina.deaconu@inria.fr\\
$^3$ Departamento de Matem\'atica Aplicada y Estad\'{\i}stica, Universidad CEU San Pablo, 28003 Madrid, Spain; lluis.hurtadogil@ceu.es\\
$^{*}$Corresponding author
}

\abstract{Recently a new algorithm for sampling posteriors of unnormalised 
probability densities, called ABC Shadow, was proposed in~\cite{StoiEtAl17}. This talk introduces a global optimisation procedure based on the 
ABC Shadow simulation dynamics. First the general method is explained, 
and then results on simulated and real data are presented. The method is 
rather general, in the sense that it applies for probability densities 
that are continuously differentiable with respect to their parameters.}


\begin{document}

\maketitle
\thispagestyle{empty}


\noindent
Let us consider that an object pattern $\yy$ is observed in a compact window $W \subset \RR^{d}$. The observed pattern is supposed to be the realisation of  a spatial process. Such a process is given by the probability density
\begin{equation}
p(\yy|\theta)=\frac{\exp[-U(\yy|\theta)]}{c(\theta)}
\label{gibbsDistribution}
\end{equation}
with $U(\yy|\theta)$ the energy function and $c(\theta)$ the normalising constant. The model given by~\eqref{gibbsDistribution} may be considered as a Gibbs process, and it may represent a Markov random field or a marked point process. Let $p(\theta|\yy)$ be the conditional distribution of the model parameters or the posterior law
\begin{equation} 
p(\theta|\yy) = \frac{\exp[-U(\yy|\theta)]p(\theta)}{Z(\yy)c(\theta)},
\label{posteriorGibbs}
\end{equation}
where $p(\theta)$ is the prior density for the model parameters and $Z(\yy)$ the normalising constant. The posterior law is defined on the parameter space $\Theta$. For simplicity, the parameter space is considered to be a compact region in $\RR^{r}$ with $r$ the size of the parameter vector.\\

\noindent
In the following, it is assumed that the probability density $p(\yy|\theta)$ is strictly positive and continuous differentiable with respect to $\theta$. This hypothesis is strong but realistic, since it is often required by practical applications.\\

\noindent
This paper proposes a Simulated Annealing (SA) method to compute~:
\begin{equation}
\widehat{\theta} = \arg \max_{\theta \in \Theta} p(\theta|\yy).
\label{maximumAPosteriori}
\end{equation}

\section{SA method~: general description}
\noindent
The SA algorithm is a global optimisation method. Assume that the probability density $p$ is to be maximised. This is achieved by sampling $p^{1/T}$ while $T \rightarrow 0$. If the temperature parameter $T$ goes to $0$ in an appropriate way, then the SA algorithm converges asymptotically towards the global optimum. This method is rather general. Under smooth assumptions, the algorithm can be generalised to minimise any criteria $U$ that can be written as $p \propto \exp(-U)$.\\

\noindent
SA algorithms for maximising probability densities for random fields and marked point process such as~\eqref{gibbsDistribution} are presented in~\cite{GemaGema84,Lies94,StoiEtAl05}. The obtained cooling schedules for the temperature parameter are of the form
\begin{equation*}
T = \frac{T_0}{1 + \log n}
\end{equation*}
with $ n > 0$. The solution guaranteed by the method converges towards the uniform distribution over the sub-space of configurations that maximises ~\eqref{gibbsDistribution}.

\noindent
The difficulty of solving~\eqref{maximumAPosteriori} is due to the fact that the normalising constant $c(\theta)$ is not available in analytic closed form. Hence, special strategies are required to sample from the posterior distribution~\eqref{posteriorGibbs}. The present paper use for thie purpose, the ABC Shadow simulation dynamics~\cite{StoiEtAl17}.\\

\noindent
The ABC Shadow dynamics is an approximate algorithm able to sample posteriors. Its main steps are presented below~:\\

\noindent
{\bf Algorithm ABC Shadow~:} fix $\Delta$ and $m$. Assume the observed pattern is $\yy$ and the current state is $\theta_0$.

\begin{enumerate}
\item Generate $\xx$ according to $p(\xx|\theta_0) = \frac{f(\xx|\theta_0)}{c(\theta_{0})}$.
\item For $k=1$ to $m$ do
\begin{itemize}  
\item Generate a new candidate $\psi$ following the density $U_\Delta(\theta_{k-1} \to \psi)$ defined by
\begin{equation}
U_\Delta (\theta \to \psi) = \frac{1}{V_\Delta} \1_{b(\theta, \Delta/2)}\{\psi\}, 
\label{uniformProposal}
\end{equation}
representing the uniform probability density over the ball $b(\theta, \Delta/2)$ of volume $V_\Delta$.
\item The new state $\theta_{k} = \psi$ is accepted with probability $\alpha_{s}(\theta_{k-1} \rightarrow \psi)$ given by
\begin{eqnarray}
\lefteqn{\alpha_{s}(\theta \rightarrow \psi) = }\nonumber \\
& = & \min\left\{1,\frac{p(\psi|\yy)}{p(\theta|\yy)}\times\frac{f(\xx | \theta)c(\psi)\1_{b(\psi,\Delta/2)}\{\theta\}}
{f (\xx | \psi)c(\theta)\1_{b(\theta,\Delta/2)}\{\psi\}}\right\} \nonumber,
\label{acceptance_probability_shadow}
\end{eqnarray}
otherwise $\theta_{k} = \theta_{k-1}$.
\end{itemize}
\item Return $\theta_m$
\item If another sample is needed, go to step $1$  with $\theta_0 = \theta_m$.
\end{enumerate}

\section{Results}
\noindent
The SA Shadow algorithm is applied here to the statistical analysis of patterns which are simulated from a Strauss model~\cite{KellRipl76,Stra75}. This model describes random patterns made of points exhibiting repulsion. Its probability density is
\begin{align}
p(\yy|\theta) & \propto \beta^{n(\yy)}\gamma^{s_{r}(\yy)} = \nonumber \\
              & =  \exp \left[ n(\yy) \log \beta  + s_r(\yy)\log \gamma \right].
\label{straussModel}
\end{align}
Here $\yy$ is a point pattern in the window $W$, while $t(\yy) = (n(\yy),s_{r}(\yy))$ and $\theta = (\log \beta, \log \gamma)$ are the sufficient statistic and the model parameter vectors, respectively. The sufficient statistics components $n(\yy)$ and $s_{r}(\yy)$ represent respectively, the number of points in $W$ and the number of pairs of points at a distance closer than $r$.\\

\noindent
The Strauss model on the unit square $W=[0,1]^2$ and with density parameters $\beta=100$, $\gamma=0.5$ and $r=0.1$, was considered. This gives for the parameter vector of the exponential model $\theta=(4.60,-0.69)$. The CFTP algorithm~(see Chapter 11 in~\cite{MollWaag04}) was used to get $1000$ samples from the model and to compute the empirical means of the sufficient statistics $\bar{t}(\yy)=(\bar{n}(\yy),\bar{s_{r}}(\yy))=(45.30,17.99)$. The SA based on the ABC Shadow algorithm was run using $\bar{t}(\yy)$ as observed data. 

\noindent
The prior density $p(\theta)$ was the uniform distribution on the interval $[0,7] \times [-7,0]$. Each time, the auxiliary variable was sampled using $100$ steps of a MH dynamics~\cite{Lies00,MollWaag04}. The $\Delta$ and $m$ parameters were set to $(0.01,0.01)$ and $200$, respectively. The algorithm was run for $10^6$ iterations. The intial temperature was set to $T_0=10^4$. For the cooling schedule a slow polynomial scheme was chosen 
\begin{equation*}
T_n = k_{T} \cdot T_{n-1}
\end{equation*}
with $k_T = 0.9999$. A similar scheme was chosen for the $\Delta$ parameters, with $k_{\Delta} = 0.99999$. Samples were kept every $10^3$ steps. This gave a total of $1000$ samples.

\begin{figure}[!htbp]
\begin{center}
\begin{tabular}{cc}
\includegraphics[width=5cm]{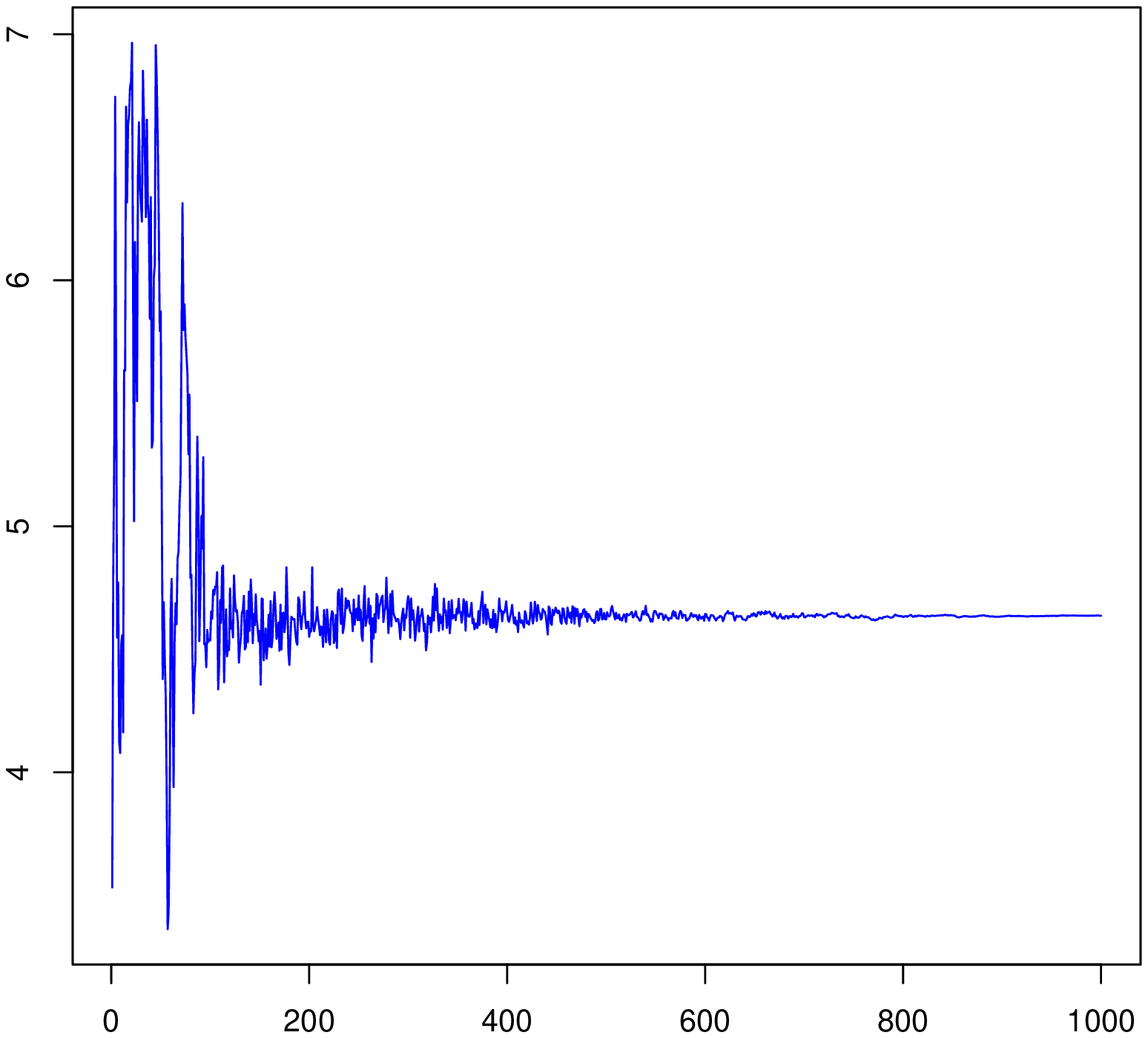} &
\includegraphics[width=5cm]{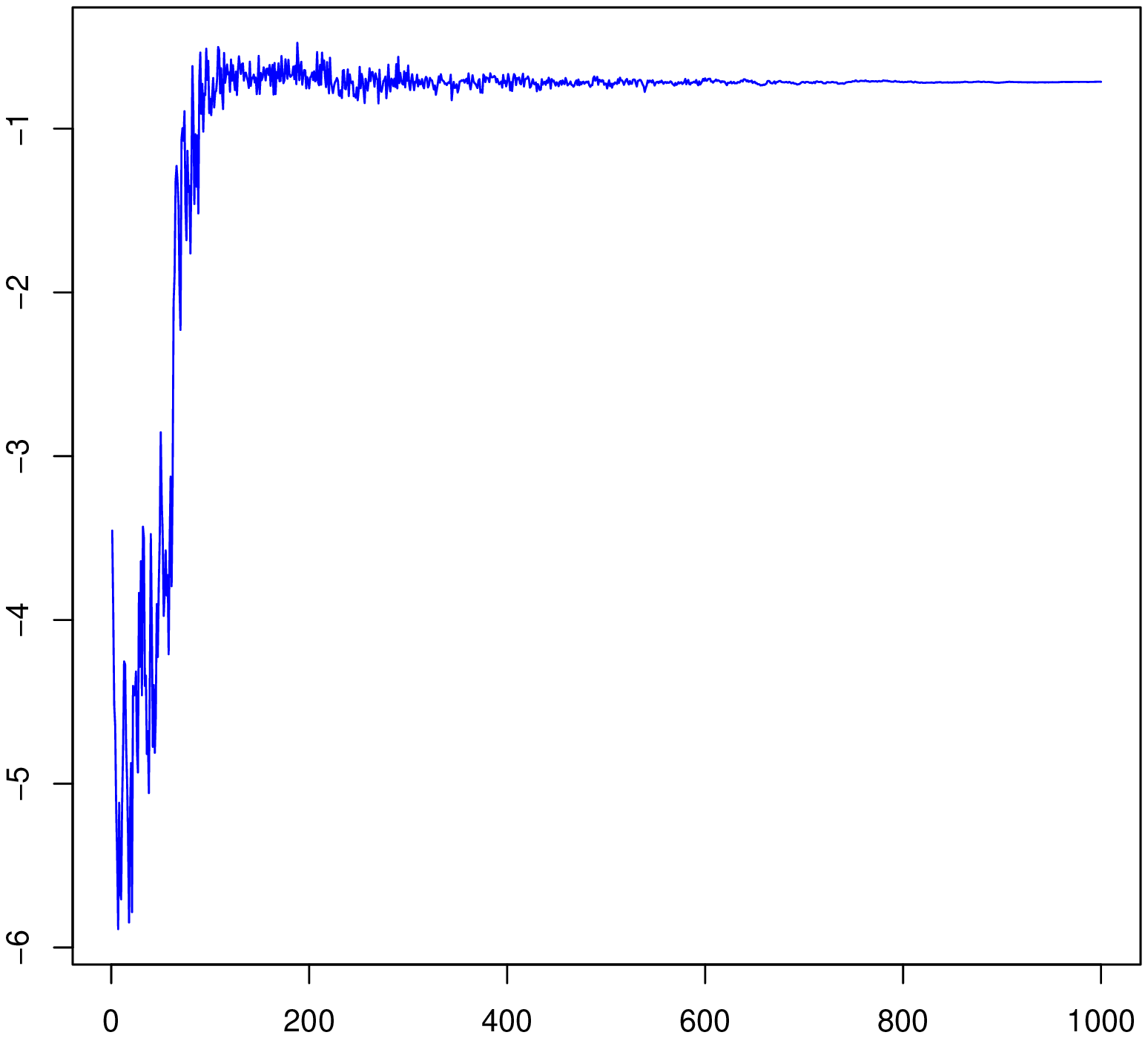}
\end{tabular}
\end{center}
\caption{SA results for computing the MAP estimates for the Strauss model parameters.}
\label{resultsStrauss}
\end{figure}

\noindent
Figure~\ref{resultsStrauss} shows the results obtained after running the SA ABC Shadow based algorithm. The final values for $\log\beta$ and $\log\gamma$ were $4.63$ and $0.71$, respectively. These values are close to the true model parameters.\\

%
%

\section{Conclusions and perspectives}
\noindent
The numerical results obtained are satisfactory. Actually, the algorithm is applied on real astronomical data, and the obtained models are tested and validated. Since the ABC Shadow is an approximate algorithm, the theoretical convergence of the SA procedure based on it, needs further investigation~\cite{HaarSaks91}.\\

\acknowledgments{Part of the work of the first author was supported by a grant of the Romanian Ministry of National Education and Scientific Research, RDI Programme for Space Technology and Avanced Research - STAR, project number 513.}

\bibliographystyle{plain}


\end{document}